\documentclass{pasj00}
\usepackage{lineno}

\begin{document}
\SetRunningHead{Mihara et al.}{MAXI GSC on ISS}

\Received{2011/02/10}
\Accepted{2011/03/22}
\Published{2011/8/25}

\title{Gas Slit Camera (GSC) onboard MAXI on ISS}

\author{
        Tatehiro \textsc{Mihara}, \altaffilmark{1}
        Motoki \textsc{Nakajima}, \altaffilmark{2}
        Mutsumi \textsc{Sugizaki}, \altaffilmark{1}
        Motoko \textsc{Serino}, \altaffilmark{1}
  	Masaru \textsc{Matsuoka}, \altaffilmark{1}
        \\
        Mitsuhiro \textsc{Kohama}, \altaffilmark{3}
        Kazuyoshi \textsc{Kawasaki}, \altaffilmark{3}
        Hiroshi \textsc{Tomida}, \altaffilmark{3}
        Shiro \textsc{Ueno}, \altaffilmark{3}
        Nobuyuki \textsc{Kawai}, \altaffilmark{4}
        \\
        Jun \textsc{Kataoka}, \altaffilmark{4}
        \thanks{Present Address: Research Institute for Science and Engineering, Waseda University, Shinjuku, Tokyo, 169-8555}
        Mikio \textsc{Morii}, \altaffilmark{4}
        Atsumasa \textsc{Yoshida}, \altaffilmark{5}
        Kazutaka \textsc{Yamaoka}, \altaffilmark{5}
        Satoshi \textsc{Nakahira}, \altaffilmark{5} 
        \\
        Hitoshi \textsc{Negoro}, \altaffilmark{6}
        Naoki \textsc{Isobe}, \altaffilmark{7}
        Makoto \textsc{Yamauchi}, \altaffilmark{8} 
        and
        Ikuya \textsc{Sakurai}, \altaffilmark{1}
        \thanks{Present Address: Research Center for Synchrotron Radiation,
          Nagoya University, Nagoya, Aichi, 464-8603}
}
\altaffiltext{1}{MAXI team, RIKEN, 2-1 Hirosawa, Wako, Saitama 351-0198}
\altaffiltext{2}{School of Dentistry at Matsudo, Nihon University, 2-870-1, Sakaecho-nishi, 
Matsudo, Chiba 271-8587}
\altaffiltext{3}{ISS Science Project Office, ISAS, JAXA, 2-1-1 Sengen, Tsukuba, Ibaraki 305-8505}
\altaffiltext{4}{Department of Physics, Tokyo Institute of Technology, 2-12-1 Ookayama, 
Meguro-ku, Tokyo 152-8551}
\altaffiltext{5}{Department of Physics and Mathematics, Aoyama Gakuin University, 
5-10-1 Fuchinobe, Sagamihara, Kanagawa 229-8558}
\altaffiltext{6}{Department of Physics, Nihon University, 1-8-14, Kanda-Surugadai, 
Chiyoda-ku, Tokyo 101-8308}
\altaffiltext{7}{Department of Astronomy, Kyoto University, Oiwake-cho, 
Sakyo-ku, Kyoto 606-8502}
\altaffiltext{8}{Department of Applied Physics, University of Miyazaki, Gakuen-Kibanadai-Nishi, 
Miyazaki, Miyazaki 889-2192}

\email{(TM) tmihara@riken.jp}





%


\maketitle

\begin{abstract}

The Gas Slit Camera (GSC) is an X-ray instrument on the MAXI (Monitor
of All-sky X-ray Image) mission on the International Space Station. It
is designed to scan the entire sky every 92-minute orbital period in
the 2--30 keV band and to achieve the highest sensitivity among the
X-ray all-sky monitors ever flown so far.
The GSC employs large-area
position-sensitive proportional counters with the total detector area of
5350 cm$^2$.  The on-board data processor has functions to format
telemetry data as well as to control the high voltage of the
proportional counters to protect them from the particle
irradiation.  The paper describes the instruments, on-board data
processing, telemetry data formats, and performance specifications
expected from the ground calibration tests.

\end{abstract}


\section{Introduction}

All-sky monitors (ASM) have been playing an important role in the
research in X-ray astronomy.  Since most X-ray sources are highly
variable, continuous monitoring of large sky fields and fast
acknowledgments of flaring events when they occur are very important
for detailed studies with follow-up observations.

The first dedicated ASM, Ariel-V \citep{Holt1976} observed a number of
X-ray novae and transients with two one-dimensional scanning pinhole
cameras in 1974--1980.  The Ginga satellite operated in 1987--1991
carried an ASM that consists of two proportional counters with six
slat collimators rotated by about 16$^\circ$ each \citep{Tsunemi1989}.
X-ray novae and transients detected by the ASM were sometimes followed
by observations with the main Large Area Counter (LAC)
\citep{Turner1989}.

The gamma-ray burst monitors such as Vela-5B \citep{Conner1969},
CGRO/BATSE \citep{Fishman1993}, and Swift/BAT \citep{Gehrels2004} also
work as ASMs.  Since 1996, RXTE/ASM has monitored X-ray intensities of
hundreds of X-ray sources as well as X-ray novae and transients
\citep{Levine1996}.  It has three cameras consisting of
one-dimensional coded masks and proportional counters.  These light
curves are open to public through the
internet\footnote{http://xte.mit.edu/ASM\_lc.html} and used by X-ray
astronomers world-wide.  The long-term data are useful, but the
detection limit is about 50 mCrab (5$\sigma$) per day.  Therefore the 
targets of RXTE/ASM are mainly on Galactic X-ray sources.

Monitor of All-sky X-ray Image (MAXI) \citep{MatsuokaMAXI} is 
a mission onboard Japanese Experimental Module - Exposed Facility (JEM-EF) on
the International Space Station (ISS).  It was
designed to achieve the better sensitivity than any ASM flown
so far using large-area proportional counters with a low background in
order to monitor long-term variability of Active Galactic Nuclei
(AGN).  MAXI carries two kinds of X-ray cameras : Gas Slit Camera
(GSC) and Solid-state Slit Camera (SSC: \cite{TsunemiSSC}; \cite{TomidaSSC}).
Both GSC
and SSC employ slit and collimator optics.  
The payload was launched by the space shuttle Endevour on July
16, 2009, then mounted on the port No.~1 on JEM-EF on July 24.
After the electric power was turned-on on August 3, MAXI started
nominal observation since August 15, 2009.  

This paper describes the design of the
GSC instrument and the expected performance from the ground
tests.  The in-orbit performance is described in \citet{Sugizaki2011}.

\section{Camera Design}



The MAXI/GSC employs the slit camera optics.  The slit camera has an
advantage of being free from the source contamination over the
coded-aperture mask while it has a disadvantage in the limited slit
area.  Thus, it is better suited for relatively faint and stable
sources such as AGNs.  To achieve the high sensitivity, large-area
proportional counters filled with Xe gas are used for the X-ray
detectors.  The total detector area of 5350 cm$^2$ using twelve gas
counters is optimized within the limit of the payload size
(0.8$\times$1.2$\times$1.8 m$^3$).



\begin{figure}
\begin{center}
  \FigureFile(85mm,){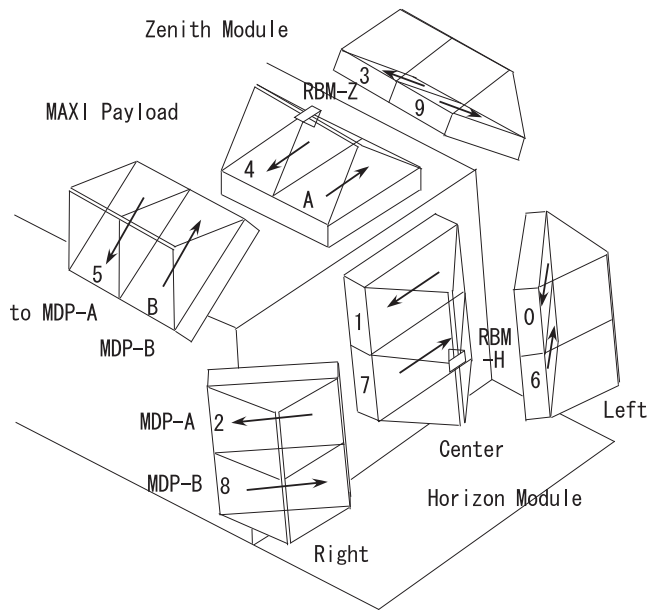} 
\end{center}
  \caption{ Schematic drawing of GSC cameras on the MAXI
    payload. The camera numbers are marked in hexadecimal as 0...9, A and B.
    Six cameras are assembled into two modules whose field of
    views are aimed to the horizontal/zenithal directions in orbit.
    Two cameras in each unit are connected to the two individual
    readout systems (MDP-A/B) independently for the redundancy.  The
    arrow in each counter indicates the X direction (DETX) of carbon
    anodes. Two RBM (Radiation-Belt Monitor) detectors are mounted 
    in the middle of the central units of 
    both the horizon and the zenith modules.
  }
  \label{fig:schematic}
\end{figure}

Figure \ref{fig:schematic} illustrates the schematic
drawing of the GSC camera units on the MAXI payload.
The entire GSC systems are composed of six identical units. Each unit
consists of a slit and slat collimator and two proportional counters
with one-dimensional position-sensitivity.  The two counters in each
unit are controlled by two individual data processors (GSC MDP-A/B in
section \ref{sec:data_processing}) via independent signal paths for
the redundancy.  

\begin{figure*}
  \begin{center}
    \FigureFile(120mm,){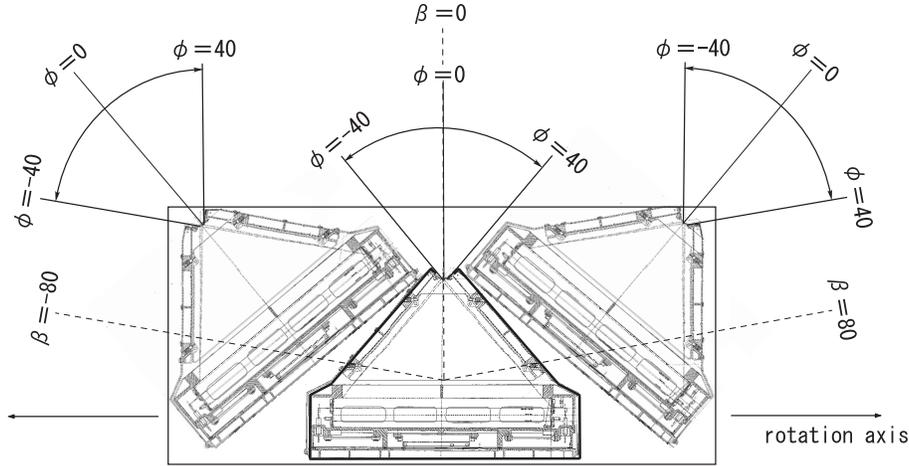} 
  \end{center}
  \caption{ Cross-section view of each horizon/zenith module.
    Three camera units are packed in a rectangular space.  The
    incident angle $\phi$ is defined for each camera, while the
    source-acquisition angle $\beta$ is defined for the MAXI
    payload. The thick outline on the center camera indicates the
    shields made of 0.1-mm lead and 0.1-mm tin sheets installed to block
    external X-ray.}
  \label{fig:3cameras}
\end{figure*}

The six camera units are assembled into two groups whose field of
views (FOVs) are pointed toward the tangential direction of the ISS
motion along the earth horizon and the zenith direction.  They are
named as horizon and zenith modules respectively.  Each horizon/zenith
module covers a wide rectangular FOV of 160$^\circ$$\times$1.5$^\circ$
(FWHM) with an almost equal geometrical area of 10--20 cm$^2$
combining three camera units.  Figure \ref{fig:3cameras} shows the
cross-section view of each module.  The areas of 10$^\circ$ from the
FOV edges to both the rotation poles are not covered because these
directions are always 
obstructed by the ISS structures.

The two FOVs of the horizon and the zenith modules 
\textcolor{black}{both scan the almost
entire sky in the 92 minutes orbital period.
Any X-ray source is, therefore,
observed twice in a orbit.}
The horizon module precedes the zenith module 
by 21.5 minutes in the normal ISS attitude.  
The FOV of the horizon unit is 
tilted up by 6 degree above the direction of the ISS motion to avoid
the earth atmosphere
as an allowance for the possible ISS attitude fluctuation.
The FOV of zenith units is set in the plane that includes the zenith 
and is perpendicular to the direction of motion.
Both FOVs have no earth
occultation and use no moving mechanics.


In the actual in-orbit operation, observation periods are limited on a
low particle-background area in order to protect the counters from the
heavy particle irradiation.  It reduces the observation duty cycle 
down to $\sim$40\% \citep{Sugizaki2011}.
The two FOVs are capable of covering the whole sky 
even if some of the counters have to be suspended from operation.
The sampling period of 92 minutes is suited to the studies of long-term
variability ($>$ 1 hour) such as AGNs.


%

%
%

\section{Detector Unit}

\subsection{Gas counter}

The GSC proportional counters employ resistive carbon fibers with a
diameter of 10 $\mu$m for anode wires.  Higher resistive anodes
are preferable for the better position resolution because the thermal
noise on the readout signal is inversely proportional to the anode
resistance.  A carbon fiber is better than a nichrome wire in this
point.  Although the traditional carbon-coated quartz has a larger
resistance, it is mechanically weak, thus easy to break by launch vibration.
The carbon fiber anodes in the
GSC were 
developed at RIKEN and were successfully used in
the HETE/WXM \citep{ShirasakiWXM}.

\begin{figure*}
    \FigureFile(85mm,){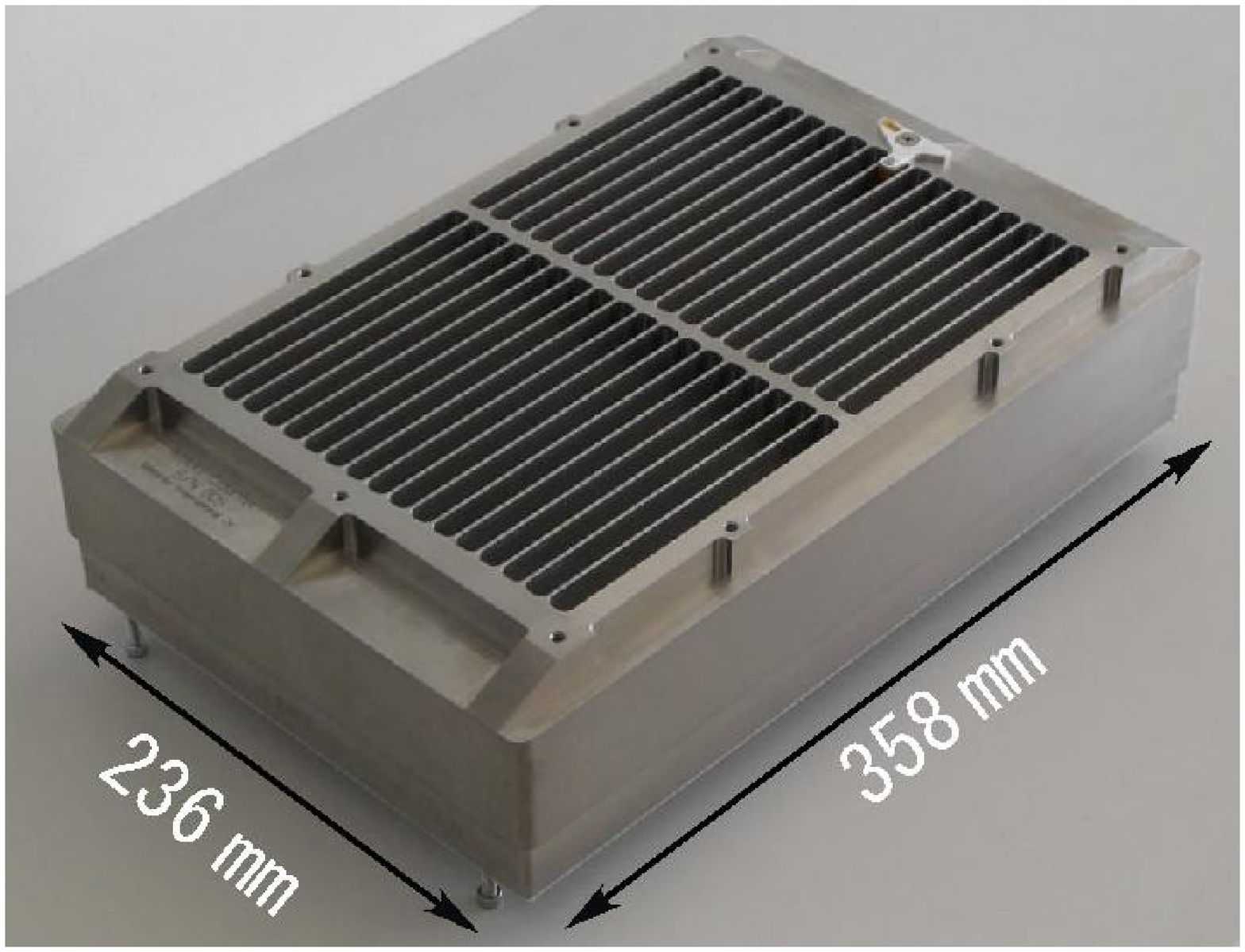} 
    \hspace{5mm}
    \FigureFile(70mm,){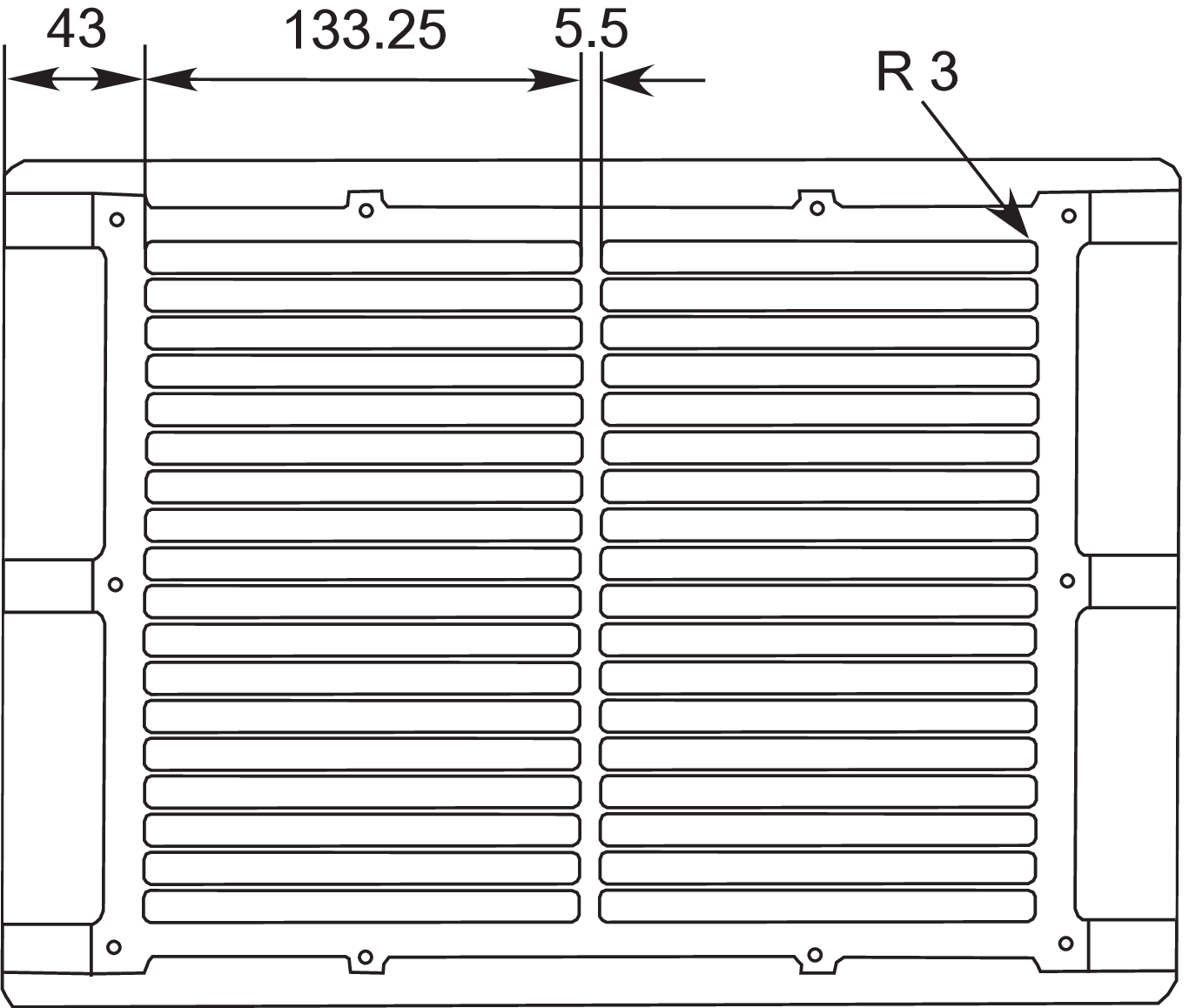} 
  \caption{ Proportional counter used in GSC 
    \textcolor{black}{
    and drawing of the window from the top.}
\textcolor{black}{
    See figure 4 and 5 for the cross-section views.
}
  . 
  }
  \label{fig:counter}
\end{figure*}

Figure \ref{fig:counter} shows a picture of a single GSC proportional
counter.  All the flight counters were manufactured by Metorex (now a
part of Oxford Instruments) in Finland.  The front X-ray window has an
area of 192 $\times$ 272 mm$^2$. It is sealed with a 100-$\mu$m-thick
beryllium foil.  To support the pressure on the beryllium foil in
vacuum that amounts 740 kgW in the whole area, grid structures with a
17-mm height are placed every 10.6-mm pitch parallel to the anode
wires.  The vertical grid is placed only at the center to keep the
open area as large as possible.  The maximum pressure of 1.66
atm is expected at the temperature of 50$^\circ$C.  
Every flight counter was tested to
withstand 1.5 times higher than the design pressure, i.e.\ 2.5 atm.
The bodies of the gas counters were made of titanium, 
which has sufficient strength and a heat expansion coefficient 
close enough to that of beryllium.
The beryllium foil is glued on the body with epoxy.
%
%

\begin{figure*}
  \begin{center}
    \FigureFile(150mm,){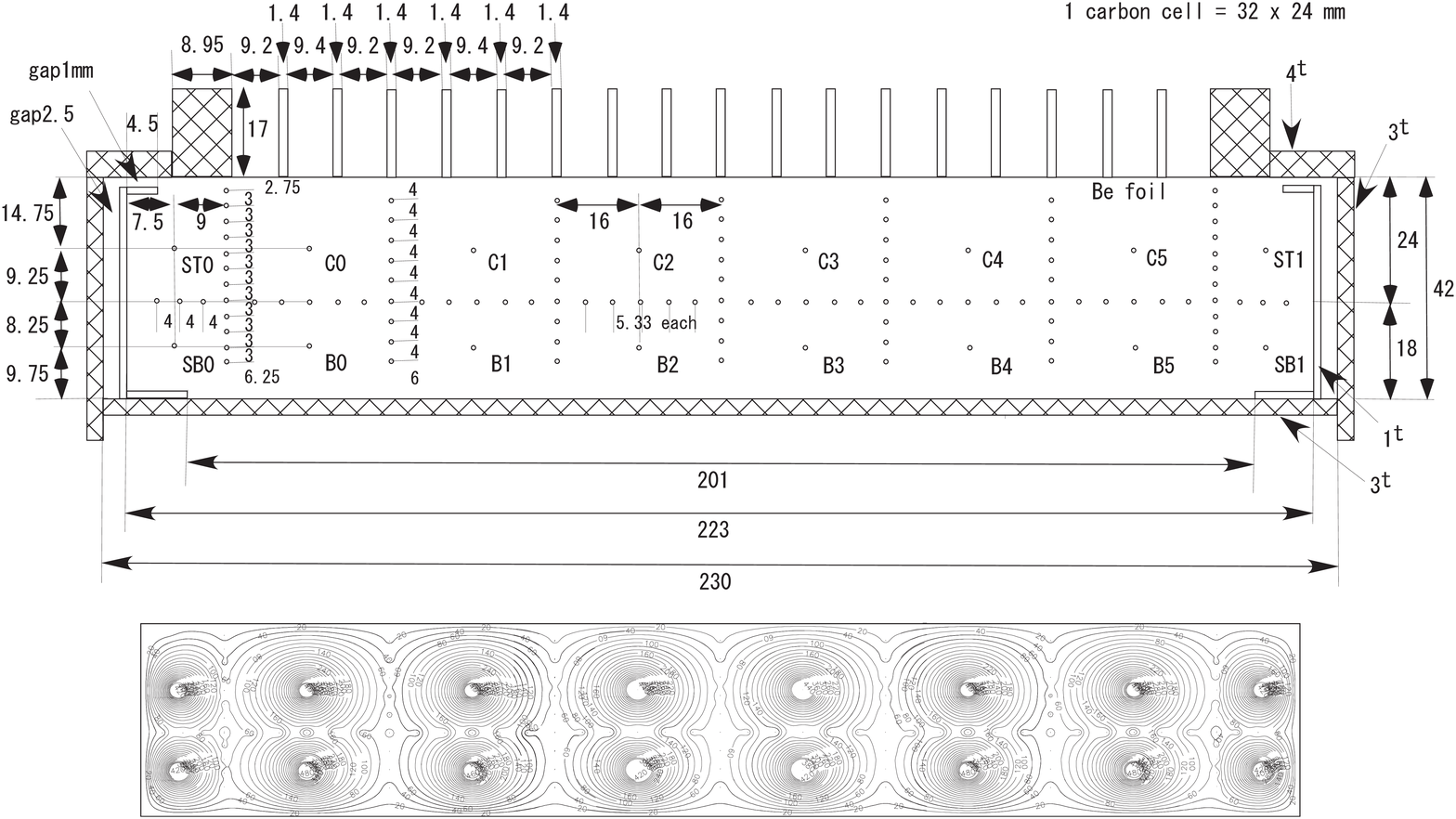} 
  \end{center}
  \caption{ Cross-section view of GSC proportional counter 
    \textcolor{black}{
    on the plane perpendicular to the anode wires.}  All
    anode/ground-wire locations are shown. The names for anode, C, B,
    ST and SB, denote Carbon, Bottom, Side Top and Side Bottom,
    respectively.  The wires of B0 to B5 are connected together in the
    counter and read out as a single bottom-veto (BV) signal.  The same for ST0,
    ST1, SB0 and SB1, as a side-veto (SV) signal. 
    All numbers represent the scale
    in units of mm. Electric potential in the counter calculated by
    Garfield is shown in the below. The ``Garfield'' is a program to
    simulate gas counters developed in CERN
    (http://garfield.web.cern.ch/garfield/).}
  \label{fig:wiregrids}
\end{figure*}

\begin{figure*}
  \begin{center}
    \FigureFile(100mm,){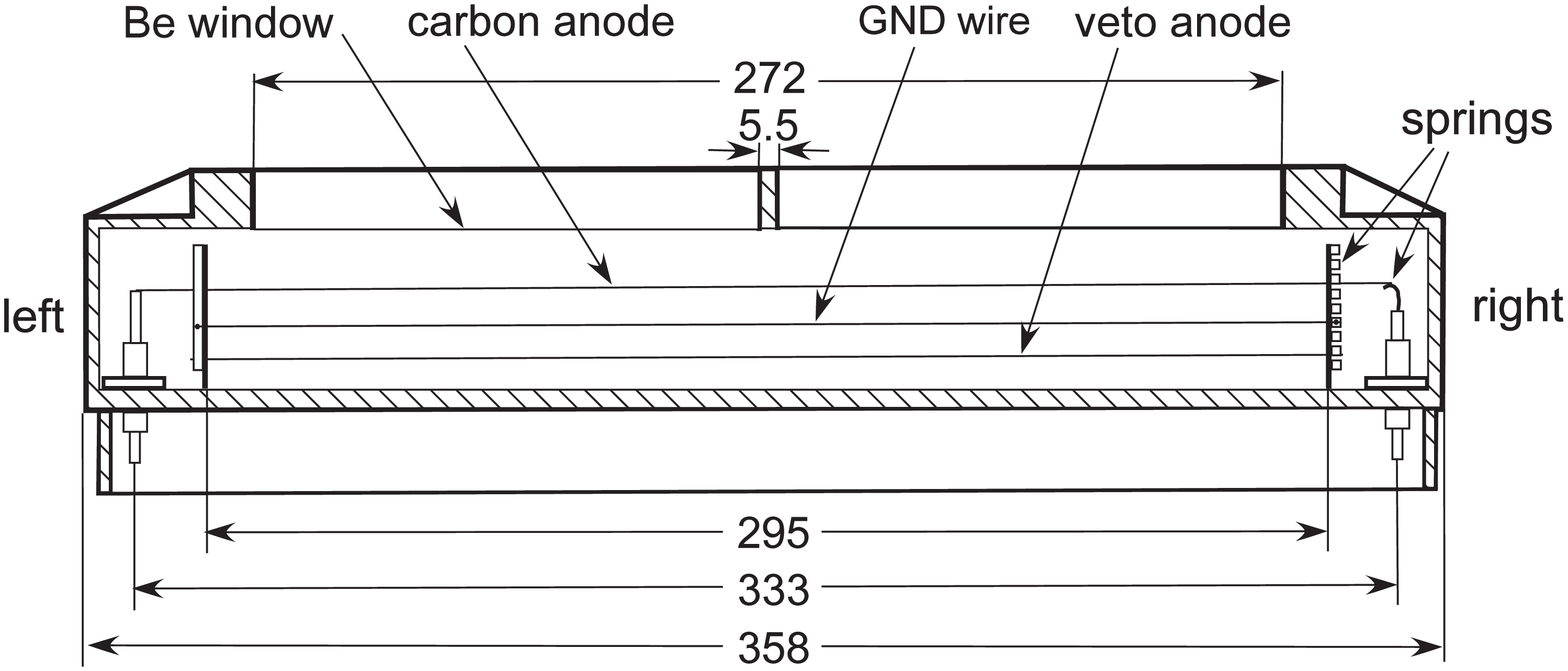} 
  \end{center}
  \caption{ Cross-section view of GSC proportional counter along the wire direction.  
    Numbers are in units of mm. }
  \label{fig:along-wire}
\end{figure*}

Figure \ref{fig:wiregrids} 
\textcolor{black}{
and \ref{fig:along-wire}}
show the counter cross-section views.  
The gas cell is divided by ground wires into six carbon-anode cells 
for X-ray detection and ten tungsten-anode cells for particle veto.
The carbon-anode layer and the bottom veto-detector layer have depths of
25 mm and 18 mm, respectively.  These sizes are determined so that the
main X-ray detectors and the veto detectors have enough efficiencies
for X-rays in the 2--30 keV band and minimum-ionization particles, 
respectively. The minimum-ionization energy in the 18-mm thick Xe
gas is 30 keV.  
\textcolor{black}{
The carbon anodes and veto anodes are not located at the center of each cell in the vertical direction.
The anode locations, 
the aspect ratios of these gas cells, and the }
spacings of the ground wires are determined so that the spatial
non-uniformity of gas gain is small within each cell.  

\textcolor{black}{
The tension of the carbon-anode wire
is set to 4 gW, which is 
sufficiently smaller than the breakage limit,
$\sim25$ gW.
All anode and ground wires are fixed via a spring at right end to 
absorb the difference of the heat expansion coefficient and keep the wires
tight and straight.}
The veto anode wires are
made of gold-coated tungsten with a 18-$\mu$m diameter,
\textcolor{black}{
which is pulled with a tension of 18 gW. We chose
as thin wires as possible for veto anodes
to achieve similarly high gas gain as the carbon anodes }
since the same high voltage (HV) is applied to both the carbon anodes and the
veto anodes. The gas-gain ratio of carbon anode to the
veto anode is 20:1.  The ground wires are made of gold-coated tungsten
with a 50 $\mu$m diameter.
\textcolor{black}{
The tension is about 50 gW.
}

We tested several kinds of gas mixture and chose a combination of
Xe (99\%) + CO$_2$ (1\%) with a pressure of 1.4 atm at 0$^\circ$C.  
The amount of CO$_2$ is decreased
from WXM PC (3\%) \citep{ShirasakiWXM} in order to reduce the spatial
gas-gain non-uniformity (section
\ref{sec:gain_spatial_non-uniformity}), and still keep the sufficient
quenching effect \citep{MiharaSPIE}.

The position resolution and the energy resolution are 
incompatible requirements.
The position resolution is primarily determined by
the thermal noise on the resistive-anode wire against the readout
signal charge.  The higher gas gain is basically preferred for the
better position resolution.  However, the high voltage for the best
position resolution is usually in the limited proportionality region
rather than the proportionality range, where the 
spatial gain non-uniformity is larger due to the space-charge effect, 
which also degrades the energy
resolution.  The operating high voltage (HV = 1650 V and the gas gain
of 7,000) 
is chosen to achieve a sufficient position resolution
and still keep an adequate energy resolution.




For the in-orbit calibration, 
a weak radioactive isotope of $^{55}$Fe is installed 
in every counter, which 
illuminates a small spot of about 1 mm in diameter
at the right end of the C2-anode cell 
(figure \ref{fig:wiregrids}).  Each isotope has a radiation of 30 kBq
and its count rate by a GSC counter is about 0.2 c s$^{-1}$ at the
launch time.

\subsection{Front-end electronics}

Each GSC counter has 
six position-sensitive anodes readout at the both ends
(left and right), and two signals for connected veto anodes.
A total 14 preamplifiers are used for the 14 analog signals.
%
%
%
%

\begin{figure*}
  \begin{center}
    \FigureFile(120mm,){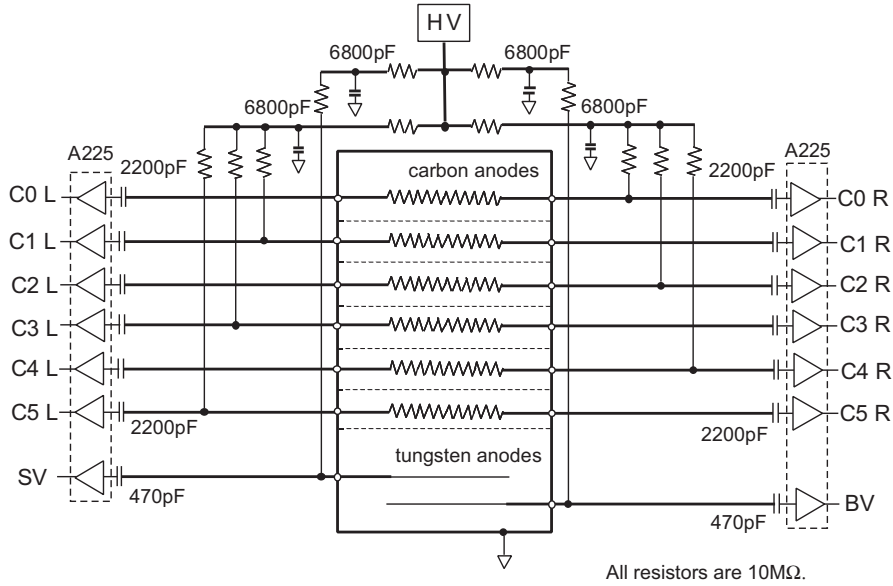} 
  \end{center}
  \caption{Schematic view of high-voltage connections to 
    anode wires on the front-end circuit board.  }
  \label{fig:HVconnection}
\end{figure*}

The front-end electronics boards are built in the backside of the
proportional counter.  It is designed to shield external noises and
also to strengthen the counter frame.  The electronics boards include
the high-voltage power supply, HK (House-Keeping) electronics and
their connectors.  The HK circuit monitors temperatures at eight
points in the camera (HV box, preamplifiers, gas cell, etc.) and HV
values. The HV-power supply with a low-power consumption was
manufactured by Meisei Electric Co.\ Ltd.  Figure
\ref{fig:HVconnection} illustrates the configuration of the HV
connections in each counter.  
\textcolor{black}{
One HV-power-supply unit works on one GSC counter.
In total, twelve HV-power-supplies are used.
}
The coupling capacitors connecting the preamplifiers and 
the anode wires are of 2200 pF, which have sufficiently lower 
impedance ($<1$ k$\Omega$ for 2$\mu$s shaping out signal of the preamplifier) 
than that of 
the carbon anodes (33 k$\Omega$) and still do not accumulate 
too much charge for the preamplifiers.

Since the wire hermetic rods come out from both the anode ends,
two front-end circuit boards are placed separately at the side ends.
We selected a hybrid-IC, Amptek A225, for the preamplifier, which is
made with a space-use quality and has a low-power consumption.

%
%

The preamplifier gains represented by the ratio of the output pulse
height to the input charge (Volt/Coulomb) should be the same between
the left and the right of each carbon anode.  We thus measured gains
of all A225 chips under the temperature condition of $-20$ to
60$^\circ$C, then selected pairs whose gains show a similar
temperature dependence.  
The feedback capacities of veto anodes are
left as they are 1 pF at the default, while those of carbon anodes are
modified to 4 pF by adding an external 3 pF capacitor
\textcolor{black}{
in order to obtain closer pulse heights for both signals from carbon anodes and veto anodes.
The ratio becomes 5:1.
}


\subsection{Slit and Slat Collimator}

\begin{figure*}
  \begin{center}
    \FigureFile(80mm,){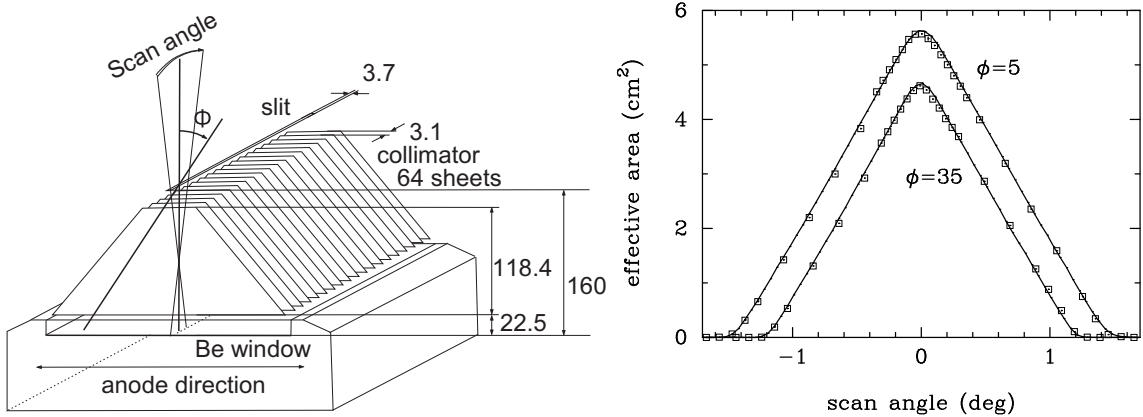}
    \FigureFile(70mm,){fig7b.eps} 
  \end{center}
  \caption{(Left): Geometry of GSC slit and slat collimator.  All
    numbers represent the scales in units of mm.  Collimator sheets
    are 0.1-mm thick and placed by 3.1-mm pitch.  (Right): Effective
    area of slit and slat collimator for $\phi = 5^\circ$,
    $35^\circ$. Points represent data measured with X-rays at Cu-K
    line (8.0 keV) in a ground test and solid lines represent the
    response models constructed for the data.}
  \label{fig:collimator}
\end{figure*}

Figure \ref{fig:collimator} illustrates the schematic view of the GSC
slit and slat collimator. The parallel tungsten rods with 3.7-mm
separation are placed at the top of the slat collimator constituting the
opening slit of the camera.  The collimator slats with a
118.4-mm height, placed at 3.1-mm pitch, constitute the FOV of
1.5$^\circ$ in FWHM, which are aligned vertically to the slit rods.
The slats are made of phosphor bronze with 0.1 mm thick.  The
thickness is determined so that X-rays up to 30 keV are stopped and
the sheets can be flattened when pulled by the tension springs.  These
surfaces are chemical-etched and roughened to avoid reflection.  64
slats are installed at the front of each counter.  The ``roofs'' of the
collimator module (as shown in a thick outline in Figure
\ref{fig:3cameras}) and the both ends of slat collimators are covered
by 0.1-mm lead and 0.1-mm tin sheets to shield Cosmic X-ray Background
(CXB). Twice thicker shield made of 0.3-mm lead and 0.1-mm tin sheets are
placed to block the direct path from the space to the beryllium
window.

The collimator transmission was tested in the JAXA beam facility.  The
setup of ground calibration is described in \citet{MoriiSPIE}.  The
transmissions are confirmed to be within $\pm5$ \% of the design.  A
model of the transmission function was constructed based on the
measured data.  Figure \ref{fig:collimator} right panel shows a
comparison between the measured data and the model.  The in-orbit
alignment calibration was carried out using celestial X-ray sources,
Sco X-1 and the Crab nebula, and is described in
\citet{MoriiCollimator}.

\section{Radiation Belt Monitor (RBM)}

The GSC is equipped with RBMs to measure the particle flux which causes
background for GSC and SSC, and to protect the instruments from
heavy irradiation.  Two RBM detectors are mounted at the
central part of both the horizon and the zenith modules
(figure \ref{fig:schematic}).


Each RBM employs a silicon PIN diode (5$\times$5 mm$^2$, 200-$\mu$m
depth) made by 
\textcolor{black}{Hamamatsu Photonics} 
for its detector.  The detector
window is covered by a 50-$\mu$m-thick aluminum whose particle
transmission is equivalent to that of the 100-$\mu$m-thick beryllium
used for the window of the main proportional counters.  The FOV is
about $80^\circ\times 50^\circ$.  The opening angle of 80$^\circ$
matches the FOV of the central camera unit.  The lower-level
discriminator (LD) can be set at 4 levels, equivalent to the deposited
energies of 30, 50, 200, and 1000 keV.  
The LD level of 30 keV is sensitive to both electron
($E > 200$ keV) and proton ($E > 2$ MeV).
The LD 50 keV is also sensitive for both electrons and
protons in the same energies, but efficiency is about 50 \% for 
\textcolor{black}{minimum-ionization particles.}
The LD 200 keV and 1000 keV are to detect only protons whose 
energies below 100 MeV and 30 MeV, respectively.
\textcolor{black}{
The energy losses of 
electrons and high energy (= minimum-ionization) protons are small 
and below 200 keV.
}

The readout electronics are designed to be fast enough to count up to
$10^5$ c s$^{-1}$, and do not saturate up to $10^7$ c s$^{-1}$.  
The count rate limit is high enough so that the RBM can tolerate 
the in-orbit rate that is expected
to be $\sim 10^4$ c s$^{-1}$ in the ISS low-altitude orbit.

The two RBMs, one on horizontal module and the other on the zenithal
module, are also considered as the redundancy.\footnote{
The results obtained in orbit show a slight difference
between the two RBMs for the anisotropy of trapped particles, which is
reported in another paper for in-orbit performance
\citep{Sugizaki2011}.
}



\section{On-board Data Processing System}
\label{sec:data_processing}

\subsection{Mission Data Processor (MDP)}
\label{sec:front_end_electronics}

\begin{figure*}
  \begin{center}
    \FigureFile(130mm,){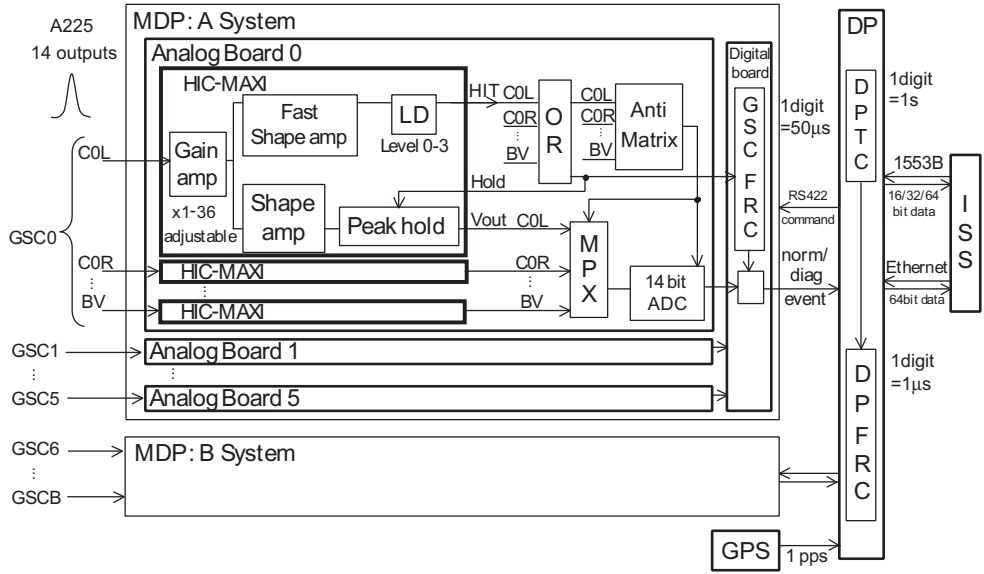} 
  \end{center}
  \caption{Block diagram of event-data readout flows on the GSC
    Mission Data Processor (MDP).


  }
  \label{fig:blockdiagram}
\end{figure*}

The GSC electronics system, named GSC Mission Data Processor (MDP,
hereafter), processes a total of 168-channel analog signals from 12
position-sensitive proportional counters.  It also have a function to
manipulates the readout event data and to receive and process commands
from the central MAXI data processor (DP).  Figure
\ref{fig:blockdiagram} illustrates the block diagram of the MDP data
processing.  The system consists of two identical modules
(MDP-A and MDP-B) for the redundancy.  Each MDP consists of six analog
boards and one digital board to process signals of six counters, a
half of the entire twelve counters.  Each analog board embodies
circuits to process signals from one counter.


We developed a hybrid IC for the GSC analog signal
processing, which is named HIC-MAXI.  
One HIC-MAXI package contains a gain amplifier, a shaping
amplifier, an LD and a peak hold circuit for a single analog channel.
Each analog board carries 14 packages.  The gain and LD are adjustable
by command to 512 levels ($\times 1.0 - 36.0$, nominal 5.0) and 4
levels (typically equivalent to the X-ray energies of 0.25, 0.5, 1.0,
2.0 keV, nominal 1.0 keV), respectively.  The design of the shaping amplifier 
is optimized for the counter position resolution.  It consists of
one-order high-pass filter with a cutoff at 145 kHz and two-order low-pass filter at
2.1 MHz.  The peak-hold capacitor of 1000 pF is chosen to sharpen the 
LD-cut threshold.  Since a MOS FET, SD215, used to discharge the peak-hold
capacitor was found to be weak for heavy irradiation, it were replaced
by the radiation-tolerant version fabricated by Calogics.
Furthermore, the HIC package was shielded with a 2-mm-thick iron.  It
then accepts the total dose requirement of 2.8 krad (for 2 years,
including the safety factor of 3).  All other electric parts have a
specification of the radiation hardness tolerant for 10 krad.


Each analog board embodies a 14-bit ADC, AD 7899, whose conversion
time is 2$\mu$s.  While 8-bit precision is sufficient for the energy resolution 
of the proportional counter, a 14-bit ADC is required for 
\textcolor{black}{
obtaining a positional value with a precision of $\sim 0.1$ mm precision
along the 333 mm anode wire even for small pulse-height events. 
}

The position along the anode wire
is encoded into the ratio of 
\textcolor{black}{two}
PHAs (Pulse Height Amplitudes) read out at the both ends.
Fourteen channels in each counter are processed by a single ADC.
Each signal is peak-held, 
then is fed to the ADC one by one through a multiplexer.

The digital processing was programmed on the fuse-type FPGAs 
operated at 10-MHz clock frequency, fabricated by ACTEL.
They are used on both the analog and digital boards.

\subsection{Event-data processing}

The front-end data processing is started when a LD-hit signal is issued at
any of 14 signal channels.  It triggers the peak-holds of all
the 14 analog outputs in each counter.  The coincidence of 
the 14 LD-hit signals is judged within the 10-$\mu$s window.  
The readout process is activated if the 
coincidence satisfies the required combination.
The hit-patterns
of the readout events can be changed by commanding.
In the nominal configuration, only those events
in which the LD-hit signal is found only from a single carbon anode are processed.

The GSC has two event-readout modes: one is the observation mode and
the other is diagnostic mode.  In the normal observation mode, two PHA
data from a carbon anode with LD hits are A/D converted, then
processed into the telemetry data.  In the diagnostic mode, all the 14
signals are processed.  The A/D conversion takes 20 $\mu$s for two PHA
data in the observation mode and 140 $\mu$s for 14 PHA data in the
diagnostic mode.  The dead time for the analog processing becomes
about 30 $\mu$s and 150 $\mu$s in observation and diagnostic modes,
respectively.  The diagnostic mode are used for the detector health
check.

The LD-hit count in each anode is always monitored by LD-hit counters.
These count data are stored in the HK data in the telemetry every second.
The counter data also includes information
of multiple hits (the numbers of carbon-carbon or veto-carbon coincidences).
\textcolor{black}{
The  hardware upper-level discriminator is not equipped on the circuit board.
Instead, the rate of events whose pulse height at either end is saturated 
in A/D conversion are monitored. 
}


\subsection{Event Timing}


Each MDP-A/B system is equipped with a 20-kHz free-run clock counter (FRC) with
a 16-bit length to measure each event time with a 50-$\mu$s precision.
When the event-readout process is activated by a LD-hit signal, the
number of the FRC is latched and included into the event data.

Each event data in the telemetry has a time tag of DP clock counter
(DPTC) that is incremented every second.  The timing of the DPTC
and FRCs on GSC MDP-A/B systems can be
calibrated on the ground using data of an on-board GPS which are also
downloaded through the telemetry.



\section{Telemetry Data}

\subsection{DP Event Data Reduction}

The MAXI Data Processor (DP) employs a multi-CPU computer system using
MIPS R3081.  The DP controls all payload instruments including GSC by
real-time and scheduled commands. It also packages the
telemetry data to be downlinked to the ground.

The DP receives GSC event data from the GSC MDP-A/B via RS-422
connection.  Each MDP can transfer 2,000 event data every second at the
maximum.  The data rate corresponds to 285 events s$^{-1}$ in the
diagnostic mode.

The event data are usually reduced in units of the 64-bit format
which consists of an camera number (4 bits) an anode ID number (4 bits), 
two pulse height data (14 bits $\times 2$)
at the both anode ends (Left and
Right), value of the 20-kHz clock counter at event timing (16 bits), hit pattern of
the 14 gas-counter signals (both ends of six carbon anodes and two
sets of veto-counter signals) (8 bits), and data-readout mode
(observation/diagnostic) (4 bits). 
The telemetry formatting occurs every second.  In the 
\textcolor{black}{
observation}
mode, each single-event data consists of the
\textcolor{black}{
64-bit-format units. In the
diagnostic mode, each event consists of seven 64-bit units;
six for the six carbon-anodes and one for the two veto anodes.
}

%
%
%


Since the telemetry bandwidth between the ISS and the ground is
limited, there are options to reduce the data mass.  The DP has a function
to filter event data with the software LD and UD, which can be set
separately for the two downlink paths (MIL-1553B and ethernet; see the next subsection), 
in each counter.  The telemetry event data also have three
reduction modes: 64-bit, 32-bit and 16-bit to be able to fit the
limited data rate.  In the 64 bit mode, the primary 64-bit-format data
are transferred without any reduction.  In the 32 bit mode, each
64-bit data are reduced to 32 bits which consists of camera number (3
bits), anode number (3 bits), two PHAs (13 bits $\times$ 2).  In the
16 bit mode, camera number (3 bits), pre-registered encoded numbers
for position (eg.\ 8 bits) and energy (eg.\ 5 bits) are returned.
The bit assignments in 16 bit mode are commandable.  
Table \ref{tab:eventformat}
summarizes the bit assignments of the three telemetry-reduction modes.
\footnote{
\textcolor{black}{
Prior to the MAXI launch 
the nominal data-reduction mode for MIL-1553B downlink was set to 
the 32-bit mode,
which continued for the initial operation phase.
It was changed to the 64-bit mode on October 30, 2009.
The data through the ethernet downlink always takes the 64 bit mode.
}
}


\begin{table*}
\begin{center}
\caption{Bit assignments of event data in three telemetry-reduction mode}
\label{tab:eventformat}
\begin{tabular}{cccccccc}
\hline
\hline
Reduction Mode    & GSCID & ANODE & PHA & GSC-FRC & HIT-PTRN & MODE\\
\hline
 64-bit  &  4 & 4 & 14$\times$2 &    16 & 8 & 4 \\
 32-bit  &  3 & 3 & 13$\times$2 &    0 & 0  & 0\\
 16-bit  &  3 & 0 & 8$^*$ (position) $+$ 5$^*$ (energy)  &    0 & 0  & 0\\
\hline
\end{tabular}
\end{center}
$^*$: The bit assignment between position and energy data are commandable within the total 13 bits. 
\end{table*}


\subsection{Data Downlink Flows and Bandwidth}

MAXI has two data downlink paths to the ground, the MIL-STD-1553B
(MIL-1553B hereafter) and the ethernet, whose maximum data-transfer
rates are 51 kbps and 600 kbps, respectively.
While the ethernet has an advantage in the speed, the MIL-1553B is superior
in the connection reliability.
GSC usually uses 50 kbps of the ethernet and SSC 200 kbps.  
SSC mainly uses the ethernet for the event data.

GSC can use about 70\% of the MIL-1553B bandwidth assigned to MAXI.  If the event rate
exceeds the limit, the overflow data are stored in the DP buffer.  The
buffered data are transferred later when the data rate becomes
lower than the limit.  The buffer for each of two readout systems (A/B)
has a volume of 500 kByte and can store 62,000 events in the 64-bit
format. The buffer data should keep the data for 50 seconds typically,
which corresponds to the single-scan duration of a point source.
Since the expected
event rate for the Crab nebula is 31 events s$^{-1}$ for each system,
bright sources with up to 40 Crab fluxes
can be observed without any data loss
in the 64bit mode. Note that the brightest X-ray source, Sco X-1, is
about 16 Crab in 2--20 keV.

%
%
%



\section{Detector protection}

The proportional counter must be operated carefully in orbit not to
suffer from any breakdown.  GSC has various protection functions in
hardware and software to avoid undesirably high count rates.
\textcolor{black}{
The MDP equips a hardware protection function using RBM counter.
The DP has four kinds of protections in software using information of RBM counts, location of ISS,
location of the sun, and the LD count rates.
}

\subsection{On-board RBM Protection}
\label{On-board RBM Protection}

A protection function using the RBM count rates is implemented in the MDP
circuit.  When either count rate of the two RBMs exceeds the
threshold, HVs of all the twelve counters are reduced to 0 V.  The
threshold of each RBM can be changed individually by commanding. The HV
suppression is released when the count becomes less than the half of
the threshold.  These judgments are made every 10 seconds.
\footnote{
\textcolor{black}{
The thresholds for both horizon and zenith RBMs are set at 100 counts in 10 seconds in the flight operation . 
The MDP-RBM flag is high during about 11 \% of the whole time, and in 3 \% of the 
low-lattitude time.}
}


\subsection{DP Software Protection}

The DP software can perform more flexible and finer controls of the
counter HVs using information collected from all the subsystems.  The DP
is always monitoring the parameters relevant to the counter
protections described in the following subsections and reduces the HVs
of the related gas counters to 0 V when any of the parameters reaches the
threshold.  The HV suppression is released automatically when they
are recognized to return to the normal condition.  The DP slowly raises
the HV not to stress the counters. The nominal rate is an increment with a 330-V step
every 10 second, which can be changed by commanding for each
counter.

\subsubsection{RBM-Count Protection}

\textcolor{black}{
The DP also offers a protection with RBM, but in a more sophisticated manner
than that by the hardware as described in section \ref{On-board RBM Protection}.
The accumulation of the RBM counts and its judgement are carried out every second
in the sliding window. The threshold level and the accumulation period (nominal 20 seconds)
are commandable.
The criteria for the HV 
reduction and the recovery are the same as the hardware one.
When either of the integrated counts of the two RBMs}
exceeds their threshold,
the HVs of all the 12 counters are reduced to 0 V.  The HV suppression
is released when both the integrated counts become less than the
halves of their thresholds.

\subsubsection{Radiation-Zone Protection} 

The Radiation-Zone (RZ) protection is designed to avoid particle
irradiation predicted from the geographical location such as SAA.  DP
keeps three RZ-map data with 2.5$^\circ$$\times$2.5$^\circ$ mesh for
three different altitudes.  When ISS enters a radiation area defined
by the RZ maps, the HVs of all the 12 counters are reduced to 0 V. The
ISS location are obtained from the ISS ancillary data, which are
broadcasted on the ISS common network of MIL-1553B.

\subsubsection{Sun Protection}

Since GSC scans the almost entire sky every orbital cycle, the
direction to the sun is also covered by some GSC camera units.  The
Sun protection is designed to avoid a direct sun illumination on the
detector, where unfavorable high count rates area expected. The
direction to the sun is calculated using the ISS ancillary data.  If
the sun is closer to the FOVs of some camera units than the limit, the
HVs of these cameras are reduced to 0 V.  The limit of the separation
between the sun direction and the camera FOV can be changed by 
commands.\footnote{
The limit of the sun angle has been reduced from 30$^\circ$ 
at the initial operation to 4$^\circ$ in the nominal operation 
to optimize the sky coverage.
}

\subsubsection{LD-Count Protection}

Every anode signal of each camera has a LD-hit counter (subsection
\ref{sec:front_end_electronics}).  The LD-count protection is designed
to avoid damages by any unexpected high count rates such as
breakdown.  Two kinds of counting methods are implemented to be
sensitive to various break modes on the detector.

\begin{enumerate}

\item Integrated High Flag\\
  If a breakdown occurs, the count-rate suddenly increases to a high
  level.  Every second, DP calculates the integrated number of LD-hit
  counts of all anodes in a camera during a given period.  If it
  exceeds the threshold, the high voltage of the counter is reduced to
  0 V.  The integration is done by the sliding window.

\item Continuous High Flag\\
  If a moderate spark occurs, the count-rate becomes slightly high
  continuously.  In such a case the continuous high flag works.  If a
  total LD-hit count of all anodes stays over the threshold for a
  given period, the high voltage of the camera is reduced to 0 V.
  
\end{enumerate}

These thresholds can be set for each counter individually by commanding.
In these LD-count protection, the HV suppression is released after a
5-minute wait time.\footnote{ 
On April 18, 2010, the ISS auxiliary data was
  stopped due to the ISS computer trouble, and the sun went into the
  FOV of a GSC camera.  The integrated high flag became on and the HV
  of the camera was shutdown safely.  At that time the sun reached
  one-fourth to the center of FOV.  The flux of the sun was measured
  1400 Crab in 2-4 keV.  
}

\section{Detector performance}

\subsection{Energy Band and Efficiency}
The efficiency curve is shown in figure 5 in \citet{MatsuokaMAXI}.  It stays
higher than 15\% in the GSC nominal energy range of 2--30 keV.  In the
lower energy end, the efficiency drops due to the absorption in the
100-$\mu$m-thick beryllium window and varies from 2 to 5\% at 1.5 keV
according to the X-ray incident angle of 0$^\circ$--40$^\circ$.  In
the higher energy range, the efficiency jumps up at the Xe K-edge
energy of 34.6 keV by a factor of 4.8.  X-ray events whose energy is
higher than the K-edge lose K-line energy due to the escape of fluorescence lines.
The escape probability of Xe K-line is 66\%, thus it has a
significant effect on the counter response.  We measured the escape
fraction and the complex energy response around the Xe K-edge energy
using GSC flight-spare counter at KEK photon 
factory\footnote{http://www.kek.jp/intra-e/research/PF.html}.


\subsection{Energy response}



The GSC gas counter is operated at the nominal HV of 1650V, which is
chosen to achieve a good position resolution.  It
is in the limited proportionality region where the energy-PHA relation
is rather distorted.  Thus, the detail calibration is necessary to
construct the energy-PHA response matrix.  The tests for the energy
response calibration were carried out for all the flight counters
using X-ray beams including fluorescence lines from various target
elements placed in the X-ray generator.  Figure \ref{fig:peaks} shows
the obtained typical energy spectra.

Figure \ref{fig:linearity} shows the energy-PHA relation derived from
the calibration-test data.  It has a discontinuity at 4.7 keV for Xe
L-edge. The relation is well reproduced by two expedient functions for
those below and above the L-edge.  The deviations between the data and
the model functions are within 0.6\% over the 2--23 keV band.

\begin{figure}
  \begin{center}
    \FigureFile(8.5cm,){fig9.eps} 
  \end{center}
  \caption{Energy spectra for K-shell emission lines from Ti (2.4 keV),
    Cr (4.5 keV), Fe (6.4 kV), Zn (8.6 keV), Se (11 keV), and Y (15
    keV).  The data were taken for pencil X-ray beams irradiated at an
    off-wire position of counter \#0 anode C0 in the nominal anode HV
    of 1650 V.  }
  \label{fig:peaks}

  \begin{center}
    \FigureFile(8.5cm,){fig10.eps} 
  \end{center}
  \caption{Linearity of energy-PHA (Pulse height amplitude) relation.
    Circles represent measured data for K-shell emission lines from
    the elements labeled.  The data were taken in the same condition
    with those of figure \ref{fig:peaks}.  Bottom panel shows residuals
    from a model expressed by expedient function, $a - b \exp(c E)$.
    The discontinuity at the Xe L-edge (4.70 keV) is implemented in
    the model.  }
  \label{fig:linearity}
\end{figure}

\begin{figure}
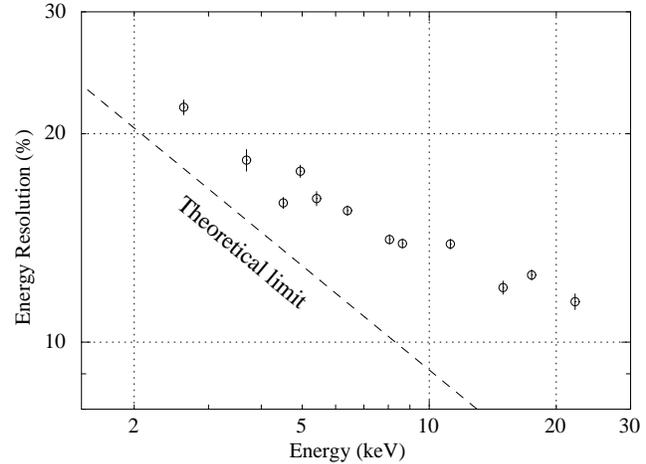

  \begin{center}
    \FigureFile(8.5cm,){fig11.eps} 
  \end{center}
    \caption{Energy resolution (FWHM) against X-ray energy. Circles with
      1-$\sigma$ error bars represent measured data with K-shell
      emission lines from the elements same as those in figure
      \ref{fig:linearity}.  The theoretical limit is estimated as
      $2.35 \{W(F+b)/E\}^{1/2}$, where the mean ionization energy $W =
      21.5$ eV, the Fano-factor $F = 0.2$, and dispersion of gas gain
      in electron avalanche $b = 0.5$ are assumed \citep{Knoll}.  }
    \label{fig:eresol}
\end{figure}

Figure \ref{fig:eresol} shows the obtained energy resolution against
the X-ray energy and that of the theoretical limit.  The measured
energy resolution is 16 \% (FWHM) at 6 keV.
It is close to the theoretical limit in the 3--5 keV
band.  The difference is larger at low and high energies.
At low energies, 
a part of the electron cloud is 
lost by the beryllium window.
At high energies, the spatial gain non-uniformity becomes effective according to
the large mean free path.

\subsection{Gain Spatial Non-uniformity}
\label{sec:gain_spatial_non-uniformity}

The gain non-uniformity along the anode wire is measured with a 2 mm pitch
using Cu K$_\alpha$ (8.0 keV)  and Mo K$_\alpha$ (17.5 keV) X-ray beams with a diameter of
0.2 mm.  The non-uniformity among each anode cell is $\sim 20$ \%
typically. The data is used in the PHA-PI conversion carried out on
the ground data reduction and also to build the energy response matrix
in the spectral analysis.

\begin{figure}
  \begin{center}
    \FigureFile(8.5cm,){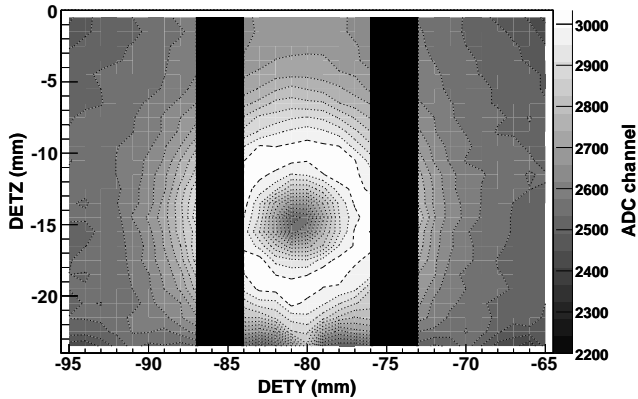}
  \end{center}
    \caption{Cross section of gas-gain non-uniformity in a single
      carbon-anode cell (C0 cell) on the plane perpendicular to the anode
      wire. The DETY direction is parallel to the detector window and
      the DETZ represent the depth from the the beryllium-window
      plane.
      The anode position is $(-80, -14.75)$.
      The data are obtained by the slant-beam method using
      X-ray beams of Mo K-line (17.5 keV).  Two black stripes are
      unmeasurable region located under the counter-support structure.
      The gray scale represents the the pulse heights of the output
      signals for the Mo K-line.  }
    \label{fig:phpres}
\end{figure}

The non-uniformity on the vertical plane to the anode wire was measured with
slant X-ray beams taking an advantage of the position sensitivity.
The method is described in \citet{MiharaSPIE}.  Figure
\ref{fig:phpres} shows the obtained gain non-uniformity around the anode
wire.  The gain is the highest in the annulus of an about 5-mm radius
from the anode. It is lower than the average at the central region
around the anode wire and the outer region.  Compared to the
distribution of the electric potential in figure \ref{fig:wiregrids},
the gain has a positive correlation with the field strength in the outer
region.

\subsection{Position Response}

Since the GSC employs slit-camera optics, the detector position
response is important for determining the direction of incident X-ray
photons with a good accuracy.  The data of the position response is
taken with a 1-mm pitch along the carbon anode using X-ray beams of Cu
K-line in the ground calibration tests.

\begin{figure}
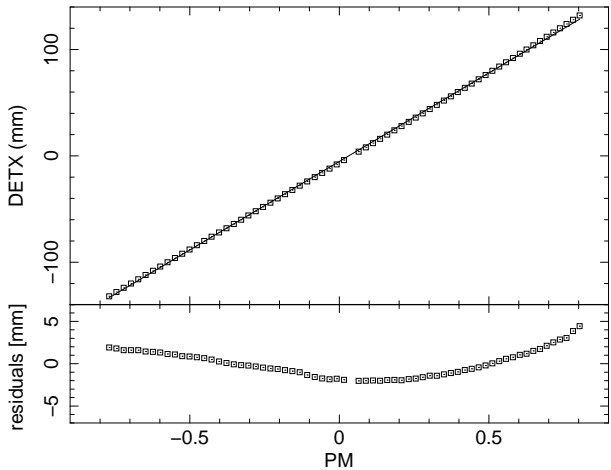

 \FigureFile(80mm,){fig13.eps} 
 \caption{\textcolor{black}{
Relation between the position at which X-ray is absorbed along the anode wire (DETX) 
and the position-measure parameter derived from the PHA ratio (PM : see text).
This is for the GSC camera 6, anode 0 measured at the nominal HV of 1650 V using pencil X-ray beam of Cu K$_\alpha$ line .
The data (square) are fitted with a linear function (solid line) 
in the top panel, and the residuals are shown in the bottom panel.
} }
 \label{scurve}
\end{figure}

The position is encoded in the ratio of the pulse height readouts at the both
anode ends.  We here define the two PHA as the left ${\it
  PHA}_{\rm L}$ and the right ${\it PHA}_{\rm R}$ and introduce a
position-measure parameter, ${\it PM}$,
\begin{equation}
{\it PM} = \frac{{\it PHA}_{\rm R}- {\it PHA}_{\rm L}}{{\it PHA}_{\rm R}+ {\it PHA}_{\rm L}}
\end{equation}
The position-measure parameter has an approximately linear relation
with the event location where the X-ray is absorbed along the anode wire
\textcolor{black}{
(figure \ref{scurve})}.
However, the relation cannot be exactly linear in the real experiment.
Any analytic function cannot successfully reproduce the data obtained
in the calibration tests with the required 
position accuracy.  We thus decided to use the table-lookup method 
in the response function based on the ground calibration tests.


\subsection{Position Resolution and High Voltage}

The position resolution should be better than the slit opening width
of 3.7mm to achieve the optimal angular resolution.  It is mostly
determined by the ratio of the thermal noise to signal charges, which are the
photoelectrons multiplied by the avalanche process in the counter.

\begin{figure}
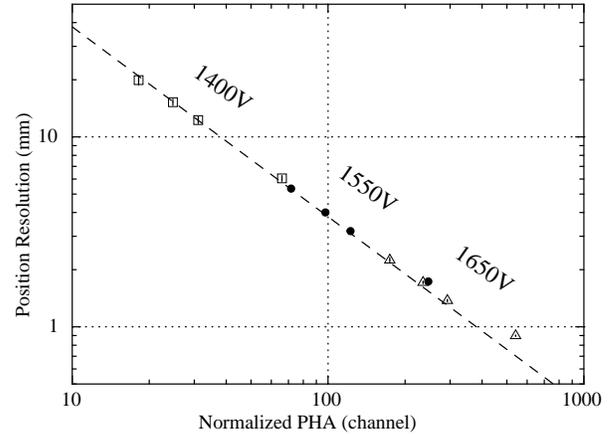

  \begin{center}
    \FigureFile(80mm,){fig14.eps} 
  \end{center}
  \caption{ Relation between position resolution (FWHM) and signal
    pulse height normalized by read-out amplifier gain.  Squares,
    filled circle, and triangle points are data taken at the HV of
    1400 V, 1550 V, and 1650 V, respectively.  Four data points at
    each HV are the resolutions for Ti-K$_\alpha$+K$_\beta$ (4.5 keV),
    Fe-K$_\alpha$ (6.4 keV), Cu-K$_\alpha$ (8.0 keV) and Mo-K$_\alpha$
    (17.5 keV). The dashed line represents an inversely proportional
    relation.  }
\label{fig:Epres}
\end{figure}

Figure \ref{fig:Epres} shows the relation between the measured position
resolution and the signal pulse height normalized by the
readout-amplifier gain, obtained from data taken for X-ray beams of
Ti-K$_\alpha$+K$_\beta$ (4.5 keV), Fe-K$_\alpha$ (6.4 keV),
Cu-K$_\alpha$ (8.0 keV) and Mo-K$_\alpha$ (17.5 keV) in three anode
voltages of 1400, 1550 and 1650 V.  It is clear that the position resolution 
is inversely proportional to the pulse height.
It is because the equivalent noise charge is constant in each anode
and the position resolution depends only on the number of the
multiplied signal charges.  The rightmost point, measured for
Mo-K$_\alpha$ at 1650 V, is slightly higher than the line because the
mean free path of the photoelectron ($\sim 0.5$ mm for 17.5 keV) is not
negligible \citep{Tabata1972}.

We determined the nominal anode voltage of 1650 V based on the
results.

\subsection{Anti-coincidence}

\begin{figure}
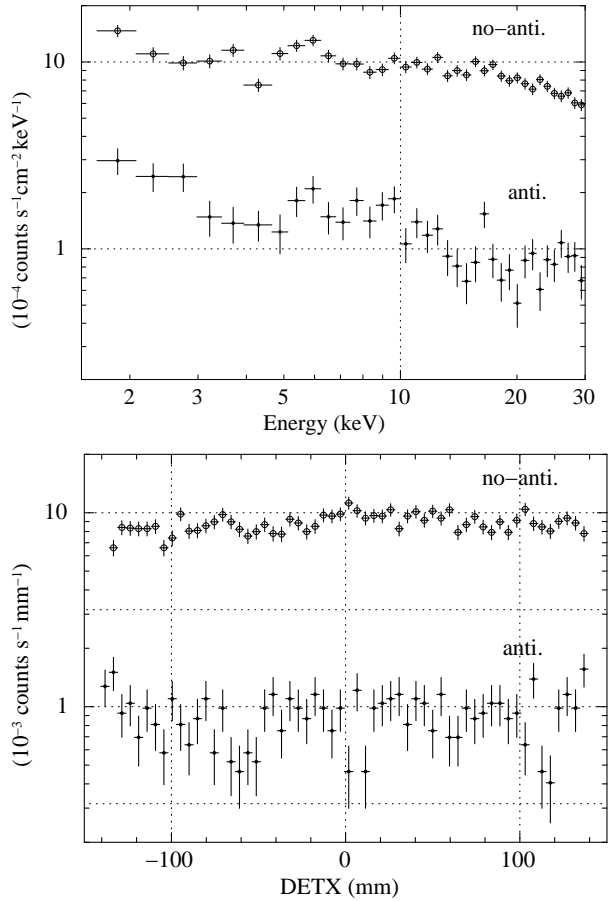

  \begin{center}
    \FigureFile(80mm,){fig15a.eps} 

    \FigureFile(80mm,){fig15b.eps} 
  \end{center}
  \caption{ Energy spectra (top) and 
    position distributions along the anode (DETX)
    (bottom) of the room background taken in the final
    ground test at Kennedy Space center in 2008 November.  The
    measurements for counter \#0 anode C0 taken with and without
    anti-coincidence cut, are shown . The anode-cell length in the
    DETX direction is $-136$ to $+136$ mm and the width is 32 mm.  }
\label{fig:antinoanti}
\end{figure}

Figure \ref{fig:antinoanti} shows energy spectra and count-rate
distributions of the room background events.
Those with and without anti-coincidence cut are shown.
The energy band of 2--30 keV is
used for the plot.
The data are obtained
as a part of the final ground tests at Kennedy Space center in 2008
November. 
The background rate at KSC was about 90 \% of that in Tsukuba Space Center.
The anti-coincidence cut reduces the event rate down to
about 1/9.  The remaining room background rate is $1.0 \times 10^{-4}$
c s$^{-1}$ cm$^{-2}$ keV$^{-1}$ in 2--30 keV.  The room background
mainly consists of gamma rays such as $^{40}$K.  The spectrum
shape does not change with and without the anti-coincidence.



\section{Summary}

The GSC is one of the X-ray instruments of the MAXI mission on the ISS. It is
designed to scan the entire sky every 92-minute orbital period in the
2--30 keV band and to achieve the highest sensitivity among the
all-sky X-ray monitors built so far. It employs large-area
position-sensitive proportional counters with a total detector area of
5350 cm$^2$ and the slit and slat collimators 
\textcolor{black}{
defining two rectangular} 
FOVs of $1.5^\circ$(FWHM)$\times 160^\circ$ to optimize the monitoring
of relatively faint sources including AGNs.
The on-board data processor (DP) has functions to format telemetry
data for the two download paths different in the bandwidth and to
protect the gas counters from particle irradiation in orbit.
The ground calibration tests confirmed the
effective area in the energy band of 2--30 keV, the energy
resolution of 16\% (FWHM) at 6 keV, the energy response in the
limited proportionality region, the spatial non-uniformity of gas gain
about 20\%, the detector position resolution of 1--4 mm varying with
X-ray energy at the nominal HV of 1650 V, and the background
rejection efficiency by the anti-coincidence technique.

\bigskip


Authors thank Metorex (Oxford Instrument), Meisei Electric Co.\ Ltd.,
and NEC Co.\ Ltd., in particular to Matti Kaipiainen for the development and
fabrication of the GSC proportional counters, Isao Tanaka, Nobuyuki
Kidachi, Koji Taguchi for the development and production of the
collimators and the electronics, and Takahiko Tanaka, Hiroshi Mondo
for the development of the DP.  
This research was partially
supported by the Ministry of Education, Culture,
Sports, Science and Technology (MEXT), Grant-in-Aid
for Science Research 19047001, 20244015, 21340043,
21740140, 22740120 and Global- COE from MEXT
"Nanoscience and Quantum Physics" and "The Next
Generation of Physics, Spun from Universality and Emergence".


\end{document}